\documentclass{aastex}

\begin{document}

\title{Spectrum Analysis of the Type Ib Supernova 1999dn:
Probable Identifications of C II and H$\alpha$}

\author{J. S. Deng \altaffilmark{1,2}, Y. L. Qiu
\altaffilmark{2} and J. Y. Hu \altaffilmark{2}}
\affil{Beijing
Astronomical Observatory, Chaoyang, Beijing 100012, P.R.China}
\email{djs@nova.bao.ac.cn, qiuyl@nova.bao.ac.cn,
hjy@class1.bao.ac.cn}

\author{K. Hatano and D. Branch}
\affil{Department of Physics and Astronomy, University of
Oklahoma, Norman, Oklahoma 73019, USA}
\email{hatano@mail.nhn.ou.edu, branch@mail.nhn.ou.edu}

\altaffiltext{1}{Corresponding author}
\altaffiltext{2}{Chinese
National Astronomical Observatories}

\begin{abstract}

Low resolution spectra of SN 1999dn at early times are presented
and compared with synthetic spectra generated with the
parameterized supernova synthetic-spectrum code SYNOW. We find
that the spectra of SN 1999dn strongly resemble those of SN 1997X
and SN 1984L, and hence we classify it as a Type Ib event.
Line-identifications are established through spectrum synthesis.
Strong evidence of both ${\rm H}{\alpha}$ and C II $\lambda 6580$
is found. We infer that ${\rm H}{\alpha}$ appears first, before
the time of maximum brightness, and then is blended with and
finally overwhelmed by the C II line after maximum; this favors a
thin high-velocity hydrogen skin in this Type Ib supernova.

\end{abstract}


\section{Introduction}

Type Ib and Ic supernovae (SNe Ib and SNe Ic) are distinguished
from SNe Ia primarily by their lack of a strong $6150{\rm~\AA}$ Si
II absorption feature in early-time spectra \citep{fil97}. Their
nature is still in debate due to their relative rareness
\citep{cap99} and faintness \citep{pie97}. The most widely
accepted scenario is the core-collapse of a massive star that has
lost its hydrogen (SN Ib) and possibly helium envelope (SN Ic)
either through a stellar wind or the effects of a close companion,
although other models like the thermonuclear explosions of white
dwarfs have not been completely excluded. Currently, there is
intense interest in SNe Ib/Ic because of their possible connection
with $\gamma$-ray bursts \citep{whe99}.

SN 1999dn, located in NGC 7714, was discovered by \citet{qiu199}
on Aug. 19.76UT with the BAO 0.6-m telescope. It was originally
identified as a Type Ic \citep{aya99,tur99} or Ib/c event
\citep{pas99}. The derived velocity of the weak ``Si II $\lambda
6355$'' absorption line is much smaller than that of other lines,
which encouraged us to reconsider the line-identification and
spectroscopic classification through spectrum synthesis. We used
the fast parameterized LTE code SYNOW and early-time spectra
observed with the BAO 2.16-m telescope. The observations, spectrum
synthesis procedure, and results are presented and discussed here.

\section{Observations}

Three spectra of SN 1999dn were obtained with the 2.16-m telescope
of Beijing Astronomical Observatory. The first spectrum was taken
with BFSOC spectrograph with a 2048x2048 pixel CCD, just after the
discovery of the supernova.  The other two spectra were taken with
the OMR spectrograph with a Tek 1024x1024 CCD.  We used a grating
of $200{\rm~\AA/mm}$, whose spectral resolution is about
$10{\rm~\AA}$. The standard IRAF (Image Reduction and Facilities)
package was used to reduce the spectra. The observations are
listed in table \ref{tbl-1}.

During the reduction process, the telluric emission lines were
subtracted automatically.  We did not remove any telluric
absorption lines from the first two spectra. We used an early-type
standard star, which has no strong features in the red part of its
spectrum, to obtain the profiles of the telluric lines and remove
them from the third supernova spectrum.  The spectra of SN 1999dn
are shown in Figures 1 --- 3.

\section{Spectrum Synthesis}

To establish line-identifications, we used the parameterized
supernova spectrum-synthesis code SYNOW to analyze the spectra of
SN 1999dn. This code makes simplifying approximations, including
the LTE, resonant scattering, and Sobolev approximations
\citep{jef90,fis97}.

\citet{hat99} have calculated LTE Sobolev line optical depths for
various temperatures and for six different compositions that could
be encountered in supernovae. We used their results to determine
which ions should be considered in our synthesis procedure. For
each ion, the optical depth of a reference line is a fitting
parameter. Optical depths of the other lines for that ion are
calculated assuming Boltzmann excitation, with excitation
temperature also a fitting parameter. The continuum is fitted with
a blackbody; its temperature $T_{bb}$ is not assigned much
physical significance.

The radial dependence of the line optical depths is taken to be
exponential with the e-folding velocity
$v_{e}=1000{\rm~km~s^{-1}}$. The velocity interval within which an ion
is present is characterized by such fitting parameters as the minimum
and maximum velocities $v_{\min}$ and $v_{\max}$. We use
$v_{\max}=5\times 10^{4}{\rm~km~s^{-1}}$, and in most cases $v_{\min}$
is the same as the velocity at the photosphere, $v_{ph}$. When we
assign to some ion a $v_{\min}$ that is larger than $v_{ph}$, we mean
that this ion is detached from the photosphere.

We have experimented with many combinations of fitting parameters.
The best-fit synthetic spectra are compared with the observed
spectra of August 21, August 31 and September 14 in the upper
panels of Figure 1, Figure 2, and Figure 3, respectively.

The synthetic spectrum of Aug. 21 has $T_{bb}=7500{\rm~K}$ and
$v_{ph}=16,000{\rm~km~s^{-1}}$. As shown in the upper panel of
Figure 1, the Fe II features from $4500{\rm~\AA}$ to
$5500{\rm~\AA}$ and the Ca II IR triplet around $8200{\rm~\AA}$
are prominent. The inconspicuous absorption near $7160{\rm~\AA}$
can also be attributed to Fe II. The absorption near
$7500{\rm~\AA}$ is usually attributed to O I $\lambda 7773$, but
the absorption here is too red for O I to fit it well. So we
introduce Mg II $\lambda\lambda 7877,7896$ and blend it with the O
I line. Both the Na I D lines and He I $\lambda 5876$ are
candidates for the strong P Cygni feature around $5700{\rm~\AA}$.
We favor He I because it can also account for other observed
features with the $\lambda\lambda 6678$ and $7065$ lines. The
absorption minimum at $6200{\rm~\AA}$ was originally identified as
Si II $\lambda 6355$ \citep{aya99,qiu299,tur99}, but through
spectrum synthesis we have found that it is better fit by the
$\lambda 6580$ line of C II, detached at $20,000{\rm~km~s^{-1}}$.
C II $\lambda\lambda 4738,4745$ may also contribute to the
absorption near $4500{\rm~\AA}$.

The observed spectrum of Aug. 31 is similar to that of Aug. 21 in
the overall shape and can be synthesized with $T_{bb}=7000{\rm~K}$
and $v_{ph}=12,000{\rm~km~s^{-1}}$. Now the C II $\lambda 6580$ P
Cygni feature becomes distinct from that of He I $\lambda 6678$.
The minimum velocity of the detached C II is now taken to be
$19,000{\rm~km~s^{-1}}$. Now the optical depth of Fe II is too
small to produce the observed absorption around $7190{\rm~\AA}$.
We find that only [O II] $\lambda\lambda 7320,7330$ can match it
while not contaminating the synthetic spectrum elsewhere. But we
are not sure that whether the physical conditions to form the
forbidden lines can be satisfied (see also \citep{fis97,mil99}).
The strong P Cygni feature near $3800{\rm~\AA}$ is Ca II H\&K.

We use the parameters $T_{bb}=5300{\rm~K}$ and
$v_{ph}=9000{\rm~km~s^{-1}}$ to synthesize the spectrum of Sep.
14. From the upper panel of Figure 3, we can see that the He I P
Cygni lines of $\lambda\lambda 5876$, $6678$ and $7065$ are
strong. He I also helps to form the observed features near
$4360{\rm~\AA}$, $4580{\rm~\AA}$ and $4860{\rm~\AA}$. The deep
absorptions at $4360{\rm~\AA}$ and $9000{\rm~\AA}$ can be matched
mostly with Mg II $\lambda 4481$ and $\lambda\lambda 9218,9244$,
which supports our identification of this ion above.  The gentle
slope from $6210{\rm~\AA}$ to $6410{\rm~\AA}$ is produced by Ca I
lines, Fe II lines, and C II $\lambda 6580$ detached at
$10,000{\rm~km~s^{-1}}$. C II $\lambda\lambda 4738,4745$, blended
with He I $\lambda 4713$, can also account for the absorption at
$4580{\rm~\AA}$. Similarly, the absorptions at $4150{\rm~\AA}$,
$5420{\rm~\AA}$ and $6130{\rm~\AA}$ can be partially or entirely
explained by Ca I. Ni II line is used to fit the absorption at
${3930 \rm~\AA}$. Note the signal-to-noise ratio per pixel around
this feature is $\sim 50$. This ion also brings two absorption
features around ${3660 \rm~\AA}$ and ${3500 \rm~\AA}$. Another
slope from $7900{\rm~\AA}$ to $8150{\rm~\AA}$ is explicable if Na
I $\lambda\lambda 8183,8195$ and C I $\lambda 8335$ are
introduced, but then the synthetic spectrum around $5700{\rm~\AA}$
and $9000{\rm~\AA}$ gets worse. For comparison, we plot the
synthetic spectrum without Ca I, Ni II, Na I and C I as the dotted
line in the upper panel of Figure 3.

The optical depths of the reference lines and other fitting
parameters for Figure 1 --- Figure 3 are listed in table 1 ---
table 3, respectively.

\section{Discussion}

\citet{pas99} indicated that the spectra of SN 1999dn were very
similar to that of SN 1997X and identified it as a Type Ib/c event
in accordance with the previous classification of the latter
\citep{gar97,mun98}, while others noted both as Type Ic supernovae
\citep{sun97,ben97,aya99,tur99}. But by comparing the spectrum of
SN 1999dn in Aug. 30 and of SN 1997X, presented by \citet{mun98},
with that of the prototypical Type Ib SN 1984L of Aug. 30, plotted
by \citet{har87} in their Figure 3, one can see that those three
spectra, all near maximum light, strikingly resemble each other
almost in all features and differ with the early-time spectrum of
Type Ic supernovae mainly in two aspects: (1) the He I optical P
Cygni lines of SN 1999dn, SN 1997X and SN 1984L are prominent,
while in the prototypical Type Ic SN 1994I He I line are hardly
discernible, although \citet{clo96} claimed evidence of very weak
He I lines in spectra with good signal-to-noise ratio; and (2)
unlike the $6150{\rm~\AA}$ absorption feature usually seen in the
early-time spectra of SNe Ic, the minimum at $\sim
6200{\rm~\AA}-6400{\rm~\AA}$ in SN 1999dn, SN 1997X and SN 1984L
is hard to attribute to Si II $\lambda 6355$, as demonstrated in
the lower panels of Figure 1 --- Figure 3 . It has been supposed
that C II $\lambda 6580$ is responsible for this feature in SNe Ib
\citep{har87}; this is supported by our spectrum synthesis here
and in particular by the identification of C II $\lambda\lambda
4738,4745$ in the Sep. 14 spectrum of SN 1999dn. We regard both SN
1999dn and SN 1997X as typical Type Ib supernovae.

The relative strengths of the He I lines $\lambda\lambda 5876$,
$6678$ and $7065$ can not be completely fitted by SYNOW.  In
Figure 1 and Figure 2, we choose to fit $\lambda 5876$ best, and
then $\lambda\lambda 6678$, $7065$ are somewhat weaker than
observed.  In Figure 3, the calculated $\lambda 6678$ is the right
strength, but $\lambda 5876$ is too strong and $\lambda 7065$ is
too weak. Such NLTE effects may be caused by nonthermal excitation
from Comptonized gamma rays released in the decay of $^{56}{\rm
Ni}$ that has been mixed into the helium envelope
\citep{shi90,luc91}. Note the Ni II line which we introduced above
to fit the ${3930 \rm~\AA}$ absorption in the spectrum of Sep. 14.
This could be direct early-time spectral evidence of $^{56}{\rm
Ni}$ mixing in the explosion of a Type Ib supernova. As for the
Aug. 31 spectrum, despite its poor signal-to-noise ratio, we are
sure that the introduction of Ni II lines to the synthetic
spectrum can not improve the fit. The velocity upper limit of the
mixed $^{56}{\rm Ni}$ may be $\sim 10,000{\rm~km~s^{-1}}$, which
is larger than the value predicted by explosion simulations
\citep{har94,kif99}.

One question remains of the minimum velocity of C II that we used
to explain the absorption near $6300{\rm~\AA}$. Why should this
ion be detached around $20,000{\rm~km~s^{-1}}$ at the earliest
times, but then detached at a much lower velocity of $\sim
10,000{\rm~km~s^{-1}}$ on Sep. 14? Does it imply two different C
II velocity components in the helium layer? An alternative
scenario that seems more attractive and probable, is that at the
earliest times this feature is produced by ${\rm H}{\alpha}$,
which always fits as well as C II $\lambda 6580$ (see the dotted
lines in the lower panels of Figure 1 --- Figure 3). In the Aug.
21 and Aug. 31 spectra, H I would be detached by
$19,000{\rm~km~s^{-1}}$ and $18,000{\rm~km~s^{-1}}$, respectively.
In the Sep. 14 spectrum, on the other hand, the identification of
C II $\lambda 6580$ is unambiguous because H I can not explain the
observed absorption at $4580{\rm~\AA}$. Actually, on Aug. 31 a
contribution from C II can not be excluded because the observed
feature is stronger than in Aug. 21 and Sep. 14. The other
synthesized H I lines like ${\rm H}{\beta}$ are swamped by Fe II
lines and are too weak to be identified.

To summarize, we suppose that a thin high-velocity hydrogen skin
exits outside of the helium layer in SN 1999dn, with ${\rm
H}{\alpha}$ appearing only before and near maximum light. As the
envelope expands and the photosphere recedes deeper into the
helium layer, the optical depth of ${\rm H}{\alpha}$ diminishes
naturally and eventually is overwhelmed by C II $\lambda 6580$. We
note that other authors \citep{fil90,jef91} have presented some
evidence, though inconclusive, of ${\rm H}{\alpha}$ in some Type
Ic supernovae, especially SN 1987M, and argued that SNe II and SNe
Ic have similar physical origins. Similarly, our identification of
${\rm H}{\alpha}$ in SN 1999dn provides a link between the
progenitors and explosion mechanisms of Type Ib supernovae and
Type IIb events such as 1987K and 1993J.

\acknowledgments

We would like to thank all observational astronomers at the 2.16m
telescope of BAO, in particular J. Y. Wei, D. W. Xu and L. Cao,
for their help in our observations of SN 1999dn. This work is
supported in part by the National Science Foundation of China.

\clearpage

\begin{deluxetable}{lcr}
\tabletypesize{\scriptsize} \tablecaption{The BAO Spectra of SN
1999dn \label{tbl-1}} \tablewidth{0pt} \tablehead{ \colhead{Date
(UT)} & \colhead{Exposure (seconds)} & \colhead{Standard}}
\startdata 1999 Aug. 21.5 & 3600 & Feige110\\ 1999 Aug. 31.6 &
2400 & BD284211\\ 1999 Sep. 14.5 & 3600 & BD284211\\
\enddata
\end{deluxetable}

\begin{deluxetable}{lcrccc}
\tabletypesize{\scriptsize} \tablecaption{Fitting Parameters for
Figure 1 ($v_{ph}=16,000{\rm~km~s^{-1}}$, $T_{bb}=7500{\rm~K}$)
\label{tbl-2}} \tablewidth{0pt} \tablehead{\colhead{Ion} &
\colhead{$\lambda({\rm\AA})$} & \colhead{$\tau$} &
\colhead{$v_{\min}(10^{3}{\rm~km~s^{-1}})$} &
\colhead{$v_{\max}(10^{3}{\rm~km~s^{-1}})$} &
\colhead{$T_{exc}(10^{3}{\rm~K})$}} \startdata He I & 5876 & 10 &
16 & 50 & 7.5\\ O I & 7772 & 1 & 16 & 50 & 7.5\\ Mg II & 4481 & 3
& 16 & 50 & 7.5\\ Ca II & 3934 & 500 & 16 & 50 & 7.5\\ Fe II &
5018 & 15 & 16 & 50 & 7.5\\ H I & 6563 & 3 & 19 & 50 & 7.5\\ C II
& 4267 & 0.02 & 20 & 50 & 7.5\\ Si II & 6347 & 3 & 16 & 50 & 7.5\\
\enddata
\end{deluxetable}

\begin{deluxetable}{lcrccc}
\tabletypesize{\scriptsize} \tablecaption{Fitting Parameters for
Figure 2 ($v_{ph}=12,000{\rm~km~s^{-1}}$, $T_{bb}=7000{\rm~K}$)
\label{tbl-3}} \tablewidth{0pt} \tablehead{\colhead{Ion} &
\colhead{$\lambda({\rm\AA})$} & \colhead{$\tau$} &
\colhead{$v_{\min}(10^{3}{\rm~km~s^{-1}})$} &
\colhead{$v_{\max}(10^{3}{\rm~km~s^{-1}})$} &
\colhead{$T_{exc}(10^{3}{\rm~K})$}} \startdata He I & 5876 & 7 &
12 & 50 & 7\\ O I & 7772 & 1 & 12 & 50 & 7\\ Mg II & 4481 & 3 & 12
& 50 & 7\\ Ca II & 3934 & 200 & 12 & 50 & 7\\ Fe II & 5018 & 10 &
12 & 50 & 7\\ $[$O II$]$ & 7321 & 0.5 & 12 & 50 & 7\\ H I & 6563 &
1 & 18 & 50 & 7\\ C II & 4267 & 0.005 & 19 & 50 & 7\\ Si II & 6347
& 2 & 12 & 50 & 7\\
\enddata
\end{deluxetable}

\begin{deluxetable}{lcrccc}
\tabletypesize{\scriptsize} \tablecaption{Fitting Parameters for
Figure 3 ($v_{ph}=9000{\rm~km~s^{-1}}$, $T_{bb}=5300{\rm~K}$)
\label{tbl-4}} \tablewidth{0pt} \tablehead{\colhead{Ion} &
\colhead{$\lambda({\rm\AA})$} & \colhead{$\tau$} &
\colhead{$v_{\min}(10^{3}{\rm~km~s^{-1}})$} &
\colhead{$v_{\max}(10^{3}{\rm~km~s^{-1}})$} &
\colhead{$T_{exc}(10^{3}{\rm~K})$}} \startdata He I & 5876 & 10 &
9 & 50 & 5.3\\ O I & 7772 & 0.3 & 9 & 50 & 5.3\\ Mg II & 4481 & 4
& 9 & 50 & 5.3\\ Ca II & 3934 & 700 & 9 & 50 & 5.3\\ Fe II & 5018
& 5 & 9 & 50 & 7\\ $[$O II$]$ & 7321 & 0.2 & 9 & 50 & 5.3\\ H I &
6563 & 0.8 & 9 & 50 & 5.3\\ C II & 4267 & 0.0005 & 10 & 50 & 5.3\\
Si II & 6347 & 0.5 & 9 & 50 & 5.3\\ C I & 9095 & 6 & 9 & 50 &
5.3\\ Na I & 5890 & 12 & 9 & 50 & 5.3\\ Ca I & 4227 & 3 & 9 & 50 &
9\\ Ni II & 4067 & 2.5 & 9 & 50 & 7\\
\enddata
\end{deluxetable}

\clearpage

\begin{figure}
\epsscale{0.8} \plotone{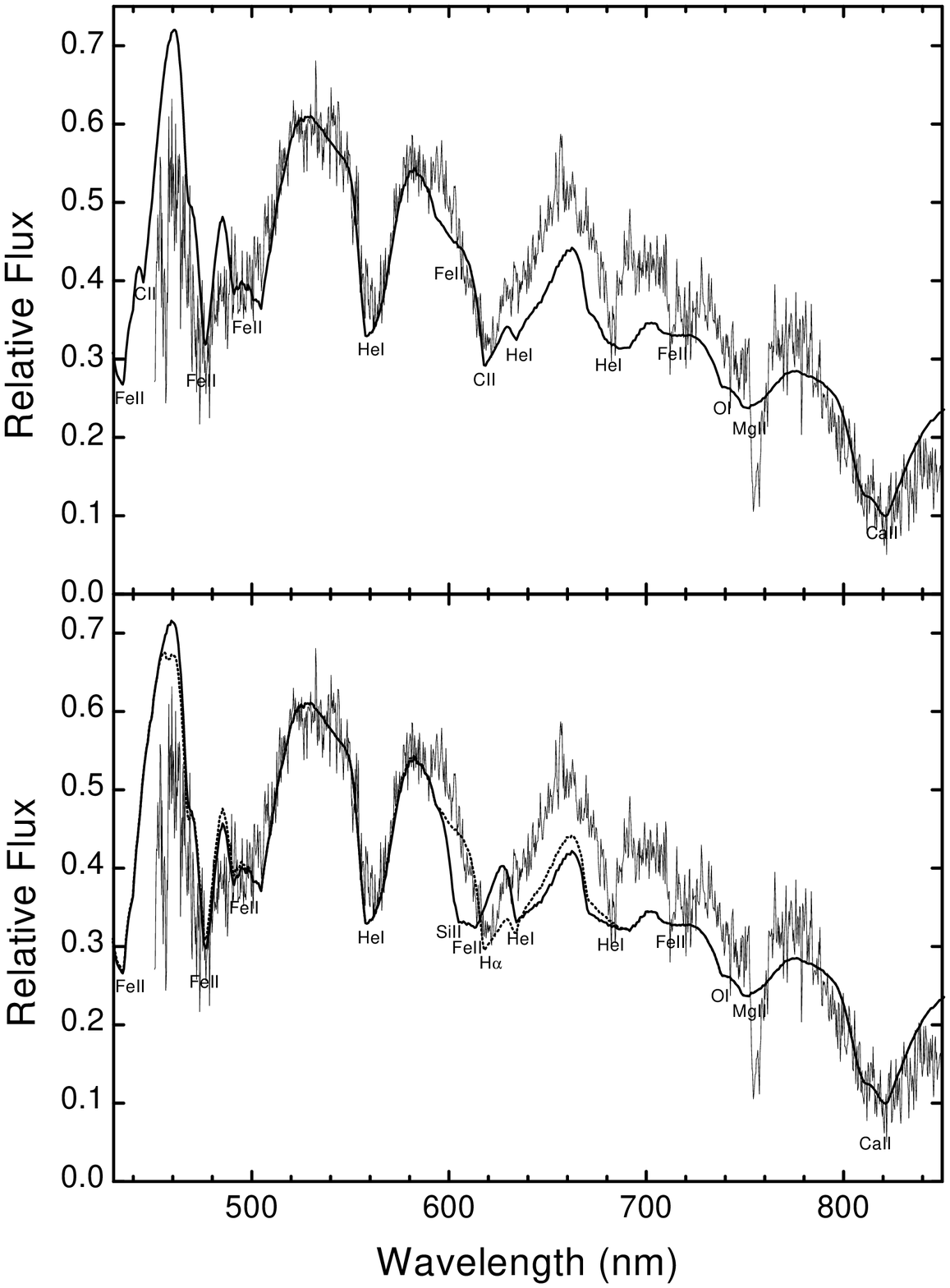} \caption{The observed spectrum of
Aug. 21 (thin solid lines) is compared with synthetic spectra
which have $T_{bb}=7500{\rm~K}$ and
$v_{ph}=16,000{\rm~km~s^{-1}}$. In the upper panel, the synthetic
spectrum with C II is plotted on the thick solid line. In the
lower panel, C II is replaced by Si II (thick solid line) and H I
(dotted line), respectively. }
\end{figure}

\begin{figure}
\epsscale{0.8} \plotone{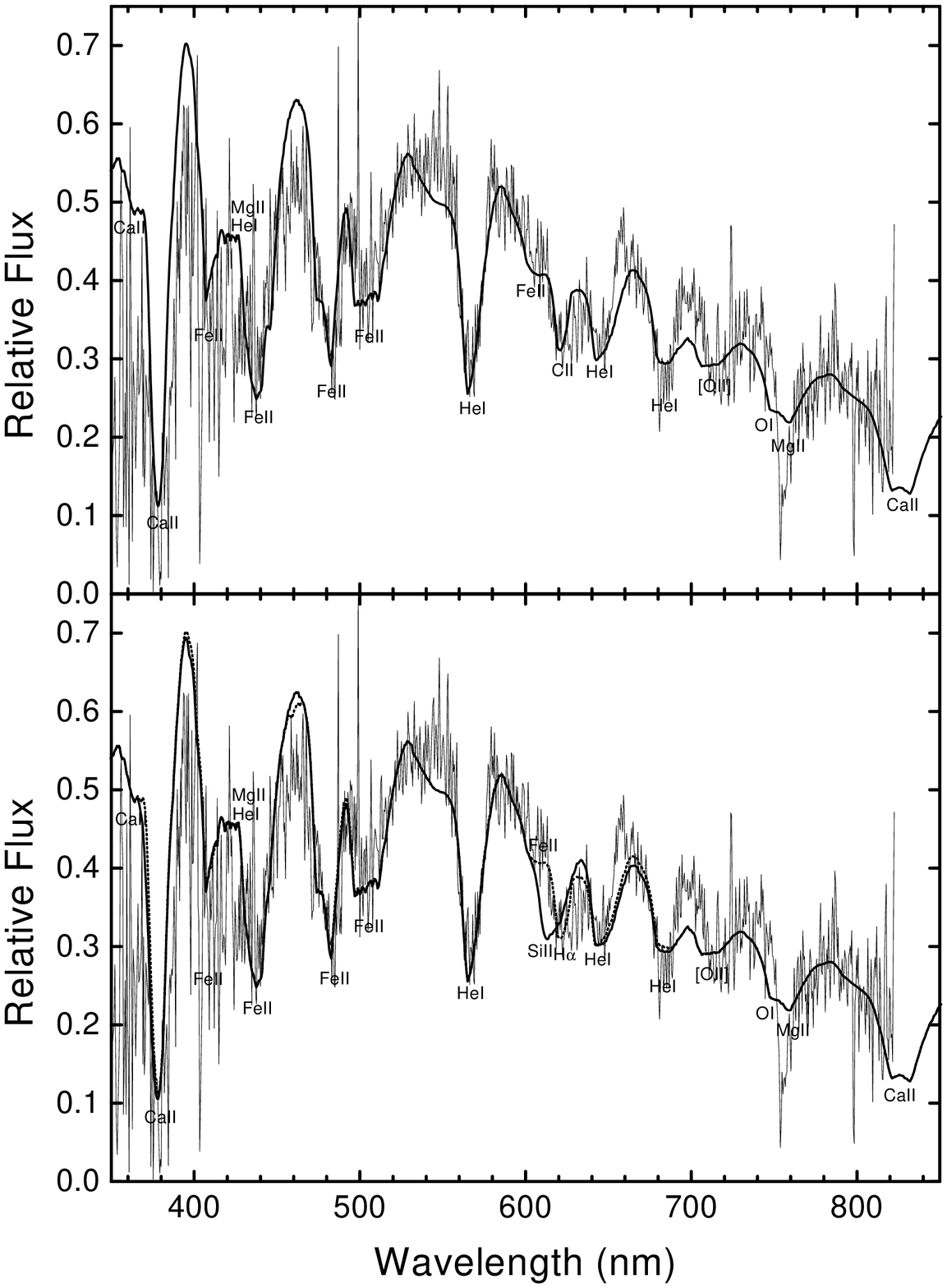} \caption{The observed spectrum of
Aug. 31 (thin solid lines) is compared with synthetic spectra
which have $T_{bb}=7000{\rm~K}$ and
$v_{ph}=12,000{\rm~km~s^{-1}}$. In the upper panel, the synthetic
spectrum with C II is plotted on the thick solid line. In the
lower panel, C II is replaced by Si II (thick solid line) and H I
(dotted line), respectively.}
\end{figure}

\begin{figure}
\epsscale{0.8} \plotone{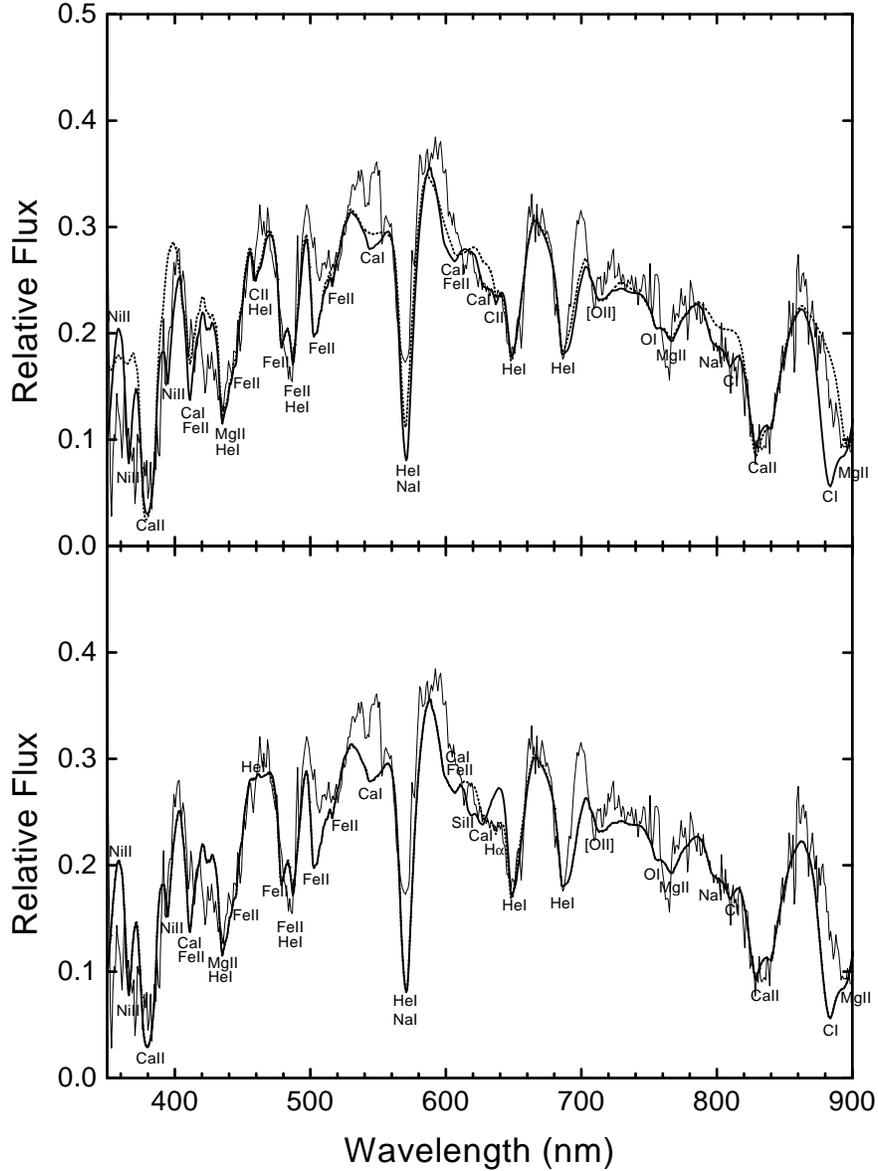} \caption{The observed spectrum of
Sep. 14 (thin solid lines) is compared with synthetic spectra
which have $T_{bb}=5300{\rm~K}$ and $v_{ph}=9000{\rm~km~s^{-1}}$.
In the upper panel, the synthetic spectrum with C II is plotted on
the thick solid line, while on the dotted line Ca I, Ni II, Na I
and C I are removed. In the lower panel, C II is replaced by Si II
(thick solid line) and H I (dotted line), respectively.}
\end{figure}

\end{document}